%% file: main.tex
\documentclass[sigconf]{acmart}

\AtBeginDocument{%
  \providecommand\BibTeX{{%
    \normalfont B\kern-0.5em{\scshape i\kern-0.25em b}\kern-0.8em\TeX}}}

\setcopyright{acmcopyright}
\copyrightyear{2022}
\acmYear{2022}
\acmConference[IDC '22]{IDC '22: Interaction Design and Children}{June 27--June 30, 2022}{Braga, Portugal}
\acmBooktitle{IDC '22: Interaction Design and Children, June 27--June 30, 2022, Braga, Portugal}
\acmPrice{15.00}
\begin{document}

\title[ARtonomous]{ARtonomous: Introducing Middle School Students to Reinforcement Learning Through Virtual Robotics}

\author{Griffin Dietz}
\affiliation{
  \institution{Stanford University}
  \city{Stanford}
  \state{CA}
  \country{USA}
}
\email{dietz@cs.stanford.edu}

\author{Jennifer King Chen}
\affiliation{
  \institution{Apple}
  \city{Cupertino}
  \state{CA}
  \country{USA}
}
\email{j_kingchen@apple.com}

\author{Jazbo Beason}
\affiliation{
  \institution{Apple}
  \city{Cupertino}
  \state{CA}
  \country{USA}
}
\email{jontait_beason@apple.com}

\author{Matthew Tarrow}
\affiliation{
  \institution{Communications by Design}
  \city{Grand Rapids}
  \state{MI}
  \country{USA}
}

\author{Adriana Hilliard}
\affiliation{
  \institution{Apple}
  \city{Cupertino}
  \state{CA}
  \country{USA}
}
\email{adri@apple.com}

\author{R. Benjamin Shapiro}
\affiliation{
  \institution{Apple}
  \city{Cupertino}
  \state{CA}
  \country{USA}
}
\email{nerd@apple.com}

% Griffin Dietz
% Jennifer King Chen
% Jazbo Beason (Ben is checking)
% Matthew Tarrow
% Adriana Hilliard
% R. Benjamin Shapiro

\renewcommand{\shortauthors}{Dietz et al.}

\begin{abstract}
\input{00-abstract}
\end{abstract}




\maketitle
\input{01-introduction}
\input{02-rw}
\input{04-system}

\input{05-method}
\input{06-results}
\input{07-discussion}
\input{08-conclusion}
\input{09-selection-of-children}

\bibliographystyle{ACM-Reference-Format}
\bibliography{references}

\end{document}

%% file: 00-abstract.tex
Typical educational robotics approaches rely on imperative programming for robot navigation. However, with the increasing presence of AI in everyday life, these approaches miss an opportunity to introduce machine learning (ML) techniques grounded in an authentic and engaging learning context. Furthermore, the needs for costly specialized equipment and ample physical space are barriers that limit access to robotics experiences for all learners. We propose ARtonomous, a relatively low-cost, virtual alternative to physical, programming-only robotics kits. With ARtonomous, students employ reinforcement learning (RL) alongside code to train and customize virtual autonomous robotic vehicles. Through a study evaluating ARtonomous, we found that middle-school students developed an understanding of RL, reported high levels of engagement, and demonstrated curiosity for learning more about ML. This research demonstrates the feasibility of an approach like ARtonomous for 1) eliminating barriers to robotics education and 2) promoting student learning and interest in RL and ML.

%% file: 01-introduction.tex
\section{Introduction}
Middle school students regularly use artificial intelligence (AI) and machine learning (ML) technologies through applications such as virtual assistants, social media algorithms, and game-playing AIs \cite{beneteau2020assumptions, keles2020systematic}. Even so, engagement with these everyday technologies does not necessarily translate to confirmed understanding---or even awareness---of AI and ML \cite{eslami2015always, eslami2016first}. Furthermore, the complexity of the technology can make it intimidating to learn and difficult to reason about \cite{beneteau2019communication}. This public knowledge gap has led researchers and educators to begin advocating for and studying AI and ML education for young learners \cite{long2020ai, touretzky2019envisioning}. In addition to supporting the development of contemporary computational literacy, AI, ML, and (closely related) data science education can foster critical engagement with data and complement existing programming-centric computing education tools and curriculum \cite{irgens2020data}. 

Robotics is currently a common entry point to computing. With over 679,000 students in 110 countries participating in FIRST robotics programs in 2019--2020 \cite{first}, some U.S. school districts (e.g., in the St. Vrain Valley of Colorado) adopting robotics programs district-wide \cite{stvrain}, and numerous robotics kits on the market (e.g., Sphero \cite{sphero}, LEGO Mindstorms \cite{lego}, and Cue Robot \cite{cuerobot}), robotics has a proven track record for engaging students in computing. However, although robotics is widely used in STEM education, the cost of specialized hardware and kits---often on the order of thousands of dollars for a classroom set---can inhibit broader reach \cite{margolis2017stuck}. Additionally, learning about robotics with today's educational robotics products typically requires purchasing ordinary computers or tablets in addition to this specialist hardware. This combination can make robotics too expensive for economically-disadvantaged communities, exacerbating inequities that already exist in computing education \cite{margolis2017stuck}.

Further, while the state-of-the-art in professional robotics makes heavy use of AI and ML (a subset of AI that learns or improves based on an algorithm), current practice in K--12 robotics education lags behind this standard, reflecting an era that predates widespread AI. Integrating ML into robotics programming can enable learners to create closed-loop, robust systems that are capable of autonomously navigating new environments without tedious manual changes.

Robotics also presents a domain that is a natural fit for exploring reinforcement learning (RL), a type of ML in which agents learn from interactions with their environment that is often leveraged in robotics contexts (see Section \ref{sec:MSML} for a detailed explanation) \cite{riedmiller2009reinforcement, sutton2018reinforcement}. Critically, this type of ML has been less studied in the AI education space than other ML approaches (e.g., supervised learning), but we conjecture that it is comparatively easy to understand. Incorporating RL into educational robotics would contribute to efforts in AI education to engage learners in rich (i.e., flexible and adaptive) and authentic (i.e., reflective of professional robotics) computational learning experiences while contributing to our understanding of K--12 education surrounding less-studied types of ML.

In our work, we leverage RL to create a robotics education experience for middle school students (ages 11--14) that increases student awareness of AI/ML, addresses barriers to wider implementation, and tests our conjecture that young learners can develop a conceptual understanding of this type of ML. Specifically, we present ARtonomous, a youth-friendly, tablet-based tool for learning about and generating reinforcement learning models for robot navigation. This system supports use with purely virtual robots, which can foster users' engagement with robotics without the startup overhead of purchasing robotics kits. Further, the autonomous navigation models are closed-loop, flexible, and abstracted, reflecting the data-driven approaches of cutting-edge robotics. These models can be used in our app in concert with user-programmable event callbacks that interface with the robot, allowing the user to customize robot behavior at specific navigational waypoints in the program. With this approach, we introduce young people to additional technical understanding and competencies for our increasingly technological world while eliminating the specialized hardware costs of physical robots.

In this paper we present the following main contributions:
\begin{enumerate}
    \item ARtonomous, a low cost, high engagement robotics education tool that helps learners explore reinforcement learning and bridges the gap between educational and professional robotics
    \item A user study evaluating how such a tool can help students develop an understanding of reinforcement learning, positively engage with the content, and build a curiosity toward machine learning more broadly
\end{enumerate}

Overall, we show that accessible and approachable reinforcement learning can bridge the gap in data-driven approaches between educational and state-of-the-art robotics. Furthermore, implementing such a system on general purpose hardware using virtual robots can reduce the cost of participation while still yielding high engagement. Our evaluation demonstrates that ARtonomous effectively introduces reinforcement learning concepts, supports student engagement, and inspires interest among middle school students in learning more about reinforcement learning and machine learning. Finally, we speculate about how the projection of virtual robots into users' physical environments using augmented reality (AR) could further enrich the impact of tools like ARtonomous.

%% file: 02-rw.tex
\section{Related Work}
Over the past several decades, there have been countless physical robotics tools ranging from modular kits to pre-constructed robots geared at introducing middle-school students to computing via a hands-on, project-based approach \cite{benitti2012exploring,blikstein2015computationally}. While these tools are successful in introducing users to programming, they do present cost challenges and very few offer access to the autonomy that is increasingly present in professional robotics technologies. Here we describe the potential for virtual robots in addressing these cost challenges, opportunities for machine learning in robotics education, and existing efforts and open avenues of inquiry in middle school machine learning education.

\begin{table*}[t]
\begin{tabular}{lccc}
\hline
& \textbf{Cost per Unit} & \textbf{Cost for Classroom Set (12)} & \textbf{Cost for 5 Sets} \\ \hline
\textbf{Makewonder Dash} & \$149.00 & \$1,788.00 & \$8,940.00 \\
\textbf{Sphero Bolt} & \$149.00 & \$1,788.00 & \$8,940.00  \\
\textbf{Sphero RVR} & \$249.00 & \$2,988.00 & \$14,940.00  \\
\textbf{LEGO Spike Prime} & \$359.95 & \$4,319.40 & \$21,597.00 \\
\textbf{LEGO Mindstorms Robot Inventor} & \$359.99 & \$4,319.88 & \$21,599.40 \\
\textbf{DJI RoboMaster S1} & \$549.00 & \$6,588.00 & \$32,940.00 \\ \hline
\end{tabular}
\caption{The price for individual and classroom sets of popular educational robotics kits in U.S. dollars. 12 units per classroom was selected based of the average middle school class size (24.9) in the United States for teachers in departmentalized instruction \cite{nces_2018} and the assumption of students working in pairs. This would be a bare minimum and require disassembly between class periods. The last column shows cost for a teacher with five classes who does not want to lose time to disassembly.}
\label{tab:robot_cost}
\end{table*}

\subsection{Addressing Cost Constraints with Virtual Robotics}
With a typical robotics kit costing in the hundreds of dollars and a classroom set in the thousands (see Table \ref{tab:robot_cost}), the cost of robotics tools can present a barrier to entry for educational institutions \cite{ching2018developing}, especially for learners of lower socio-economic status. These robotics tools also usually need to be accompanied by tablets or computers used to write the code to control the bots and enough dedicated physical space in which to operate them \cite{first_2021}, further increasing barriers to implementing robotics tools into school curriculum. 

Prior work demonstrates that educational robotics tools need not necessarily be expensive. Some research has shown how very motivated learners with expert teachers can re-purpose surplus technology to support robotics education \cite{blikstein2008travels}. Further, lower priced, maker-oriented robotics tools can also provide entry points to robotics (e.g., Sparkfun's micro:bot system), though with fewer capabilities than more expensive products. Nonetheless, the high cost of common educational robotics products leads us to wonder if new robotics education technologies that heavily leverage the capabilities of general purpose computing devices \cite{bellas2017robobo} can lower economic barriers to entry while cultivating interest and knowledge development that could scaffold \cite{pea2004social} deepened participation over time.

Virtual approaches for educational robotics remove the need for costly peripherals on top of the required computer or tablet, thereby making robotics more accessible while still successfully engaging students with key ideas in STEM \cite{berland2015comparing,eguchi2012student,fernandes2019teaching,witherspoon2017developing}. AR functionality has the potential to take this engagement and learning even further \cite{dede2009immersive, cheli2018towards}, yet AR-robot integration to date focuses on entertainment \cite{mariokart}, communication \cite{villanueva2021robotar, walker2018communicating}, or debugging features for physical robots \cite{cao2019v, akan2011intuitive}, rather than on hardware alternatives for education. We therefore aim to lower the cost barrier to educational robotics while still encouraging high engagement by incorporating virtual robots driven by general purpose computing devices.
% We believe that virtual approaches can lower the cost barrier to educational robotics while still encouraging high engagement.

\subsection{Autonomy in Robotics Education}
Typical robotics kits utilize motor-powered wheels for movement (e.g., mBot \cite{mbot} and LEGO Mindstorms Robot Inventor \cite{lego}). These systems allow users to specify specific amounts of time or distances to move, but using these products on different surfaces (e.g., carpet or tile) can lead to substantially different distances actually traveled due to varying degrees of friction. Users can address these inconsistencies by modifying their code for use in different environments, but this approach allows for only short-term fixes and remains inflexible to changes in route. 

Critically, this environmental rigidity is inconsistent with state-of-the-art robotic systems that automatically adapt to changing environments, presenting an opportunity for the introduction of machine learning techniques. To date, navigation of unknown settings in educational robotics relies on information gathered from sensors, but few educational robotics tools enable learners to create autonomous navigation control systems. The DJI RobotMaster S1 uses an on-board camera to acquire location information (e.g., distance from a potential barrier) and will continue to move unless within a specified distance of an obstacle. The S1 can then rotate, attempting to locate a safe direction to head, although this decision-making is still contingent on user programming \cite{s1}.

We envision new educational tools that empower learners to create even more autonomous systems, with projects that get from point A to point B without relying on fine-grained, route-specific programming of the steps in between. Robotics tools that enable novices to build autonomous navigation systems with ML can increase the resiliency and authenticity of what they can create, support AI education (e.g., about RL), and offer rich contexts to integrate ML with programming (such as to respond dynamically to events that arise during a robot's autonomous navigation). 

\subsection{Middle School Machine Learning Education} \label{sec:MSML}
In recent years, researchers have begun to explore introducing K--12 students to machine learning topics and to develop curricular objectives to this end \cite{touretzky2019envisioning, long2020ai}. From these K--12 AI literacy objectives, we can identify the importance of supporting learners in building a mental model of the algorithm's inputs, representations, and decision-making processes and understanding what learning means for a machine, including the role of data and of the human \cite{touretzky2019envisioning, long2020ai}.

Existing machine learning education tools for K--12 tend to focus on the use \cite{payne2021danceon, AppInventorAICurriculum} or construction \cite{agassi2019scratch, druga2018growing, kahn2017child, zimmermann2019youth, zimmermann2020youth} of classifiers. In cases where students create their own classifiers, they typically do so via supervised learning, and once created, these classifiers can be used within code in common block-based programming environments \cite{agassi2019scratch, druga2018growing, kahn2017child, zimmermann2020youth} or operate independently from programming tools altogether as standalone tools for gesture recognition \cite{zimmermann2019youth}. While using these systems, some learners enacted engineering practices that distinctly encapsulated modeling and programming practices: separately writing code and training models, then independently testing each type of product, before integrating their models and code within their Scratch games \cite{zimmermann2020youth}. Those results inform a key aim of the present system and study: creating new tools that can help students learn to recognize and apply the complementary capabilities of ML and programming within robotics systems. 

Few tools for K--12 AI education exist that go beyond using or creating supervised learning classifiers, although those that do explore the teaching of k-means clustering, a form of unsupervised learning \cite{wan2020smileycluster} or generative adversarial networks (GANs) \cite{ali2021exploring}. 

Reinforcement learning (RL) is another genre of machine learning in which computational agents learn from interactions with their environment \cite{sutton2018reinforcement}. Like in supervised learning, RL agents receive feedback as they learn. However, the \textit{kind} of feedback they receive differs. In supervised learning, ML algorithms are provided with correct information about what they should have predicted, typically coming from a human-annotated dataset. In RL, the learning algorithm still receives feedback, but rather than gaining information about what the system's correct action (i.e., prediction) should have been, the system receives only a positive or negative \textit{reward} that it uses to inform future decisions. 

Critically, RL is frequently used in research on autonomous systems, including in robotics \cite{riedmiller2009reinforcement}, self-driving cars \cite{sallab2017deep}, and game-playing AIs  \cite{silver2018general}. At every step, these agents choose an action to take from the action space given the current input, receive some reward or penalty from their environment based on that action, and learn how to take better actions in the future based on the reward/penalty that they received \cite{sutton2018reinforcement}.

Despite the importance of robotics to middle school computing education and the role of RL in state-of-the-art robotics, little research to date has described tools and methods for introducing RL to middle school students, in the robotics context or otherwise, although some such systems exist for adult users \cite{law2021hammers, martinez2019teaching}. Notably, Zhang et al. recently published an RL robotics system that shares in our goal of helping students better understand and train robots \cite{zhang2021interactive, zhang2021augmented}. This work includes a pilot assessment of student engagement and learning that demonstrates satisfactory learning outcomes \cite{zhang2021interactive}. However, unlike our platform, this tool focuses on training physical robots (LEGO Spike Prime; see Table \ref{tab:robot_cost} for cost) and targets high school students as users \cite{zhang2021interactive}.

We view the training of robotic control systems and the integration of those RL models with code as essential practices for future robotics education. Therefore, building on prior work, we add RL to middle school students' robotics construction toolbox in a manner that develops students' understanding of RL, inspires engagement, and motivates curiosity to learn more.

%% file: 04-system.tex
\section{System Design \& Implementation}
\begin{figure*}[t]
  \centering
  \includegraphics[width=\linewidth]{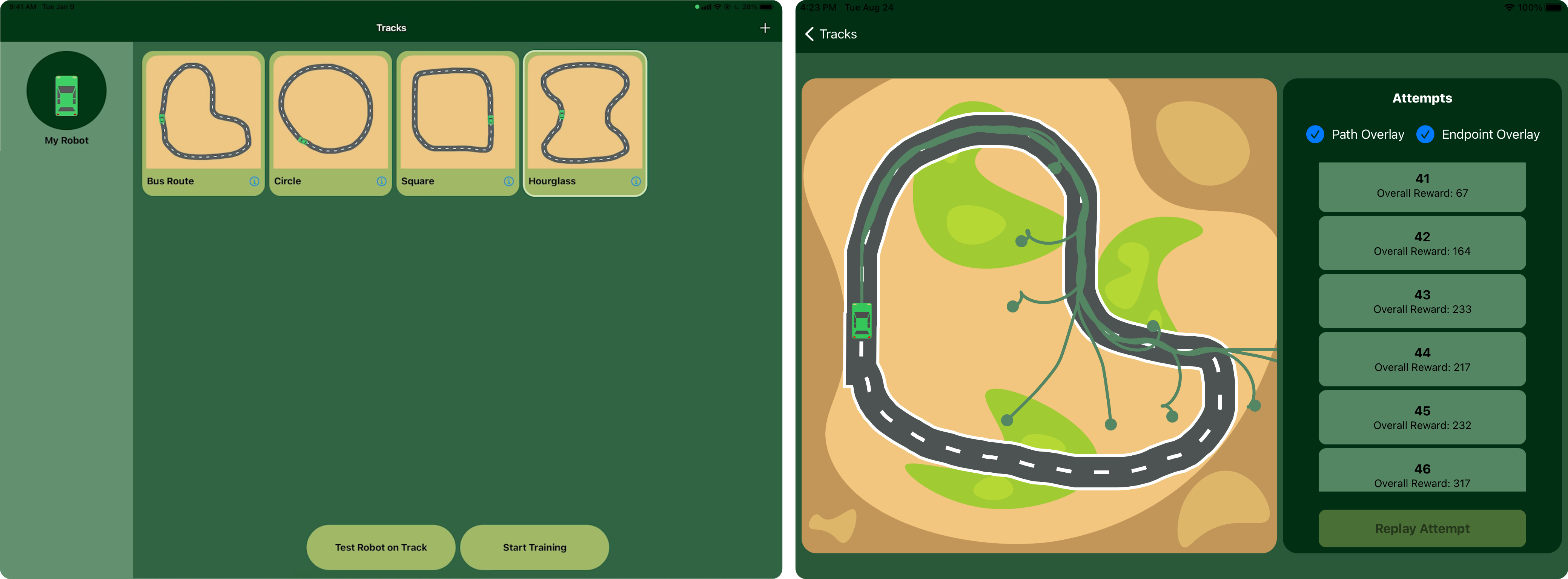}
  \caption{\textit{Left: }From the track dashboard, students can create new tracks, test on an existing track, train on an existing track, or select a track to view training details. \textit{Right: }On the training detail page users can view overlays showing the paths and endpoints of each episode on the selected track, along with a list of episodes and their overall reward. Users can select individual episodes to replay.}
  \label{fig:training-detail}
\end{figure*}
We developed ARtonomous, an iPad application to bring RL into middle school robotics. This two-part learning tool allows students to first train RL models for autonomous virtual robot navigation and then write imperative code that interfaces with these models to handle specific events.

\subsection{Design and Learning Objectives}
Based on prior work and our aforementioned research objectives, we aimed to create a low-cost, high engagement robotics experience that helps learners explore AI/ML. Our system design therefore needed to 1) leverage virtual robots to run entirely on a general-purpose device (i.e., an iPad) and eliminate the cost of peripherals and 2) support the creation of RL autonomous navigation models to introduce key ideas in machine learning. Additionally, we aimed to make it easy to get started with this autonomous navigation approach by creating introductory teaching materials that we used within this study to orient participants.

Similarly, based on prior ML education systems research \cite{zimmermann2019youth,druga2018growing} and curricular frameworks \cite{touretzky2019envisioning, long2020ai}, we aimed to support learning objectives pertaining to developing understanding of and curiosity for RL. Specifically, students should:
\begin{itemize}
    \item develop a fundamental understanding of RL (e.g., understanding episodes, rewards, and penalties) while also engaging with coding fundamentals
    \item use a combination of quantitative performance metrics and qualitative observations to find problems with individual models and generate ideas for how to fix them
    \item be able to identify which problems in robotics might be solved with code and which might be solved with RL
    \item use their experiences to develop a broader understanding or curiosity around RL and ML systems outside robotics
\end{itemize}

\subsubsection{From Vision to Prototype}
In our present work, we created a system to investigate the feasibility, usability, and excitement generated by the core ideas of these design and learning objectives: individual students use a tablet to trace a route for a robot, iteratively train an RL model for a virtual robot to navigate that route, and then write code to control how the virtual robot behaves at waypoints along its route. The complete system has two components, the ARtonomous model training application and the ARtonomous programming tool.

\begin{figure*}[t]
  \centering
  \includegraphics[width=\linewidth]{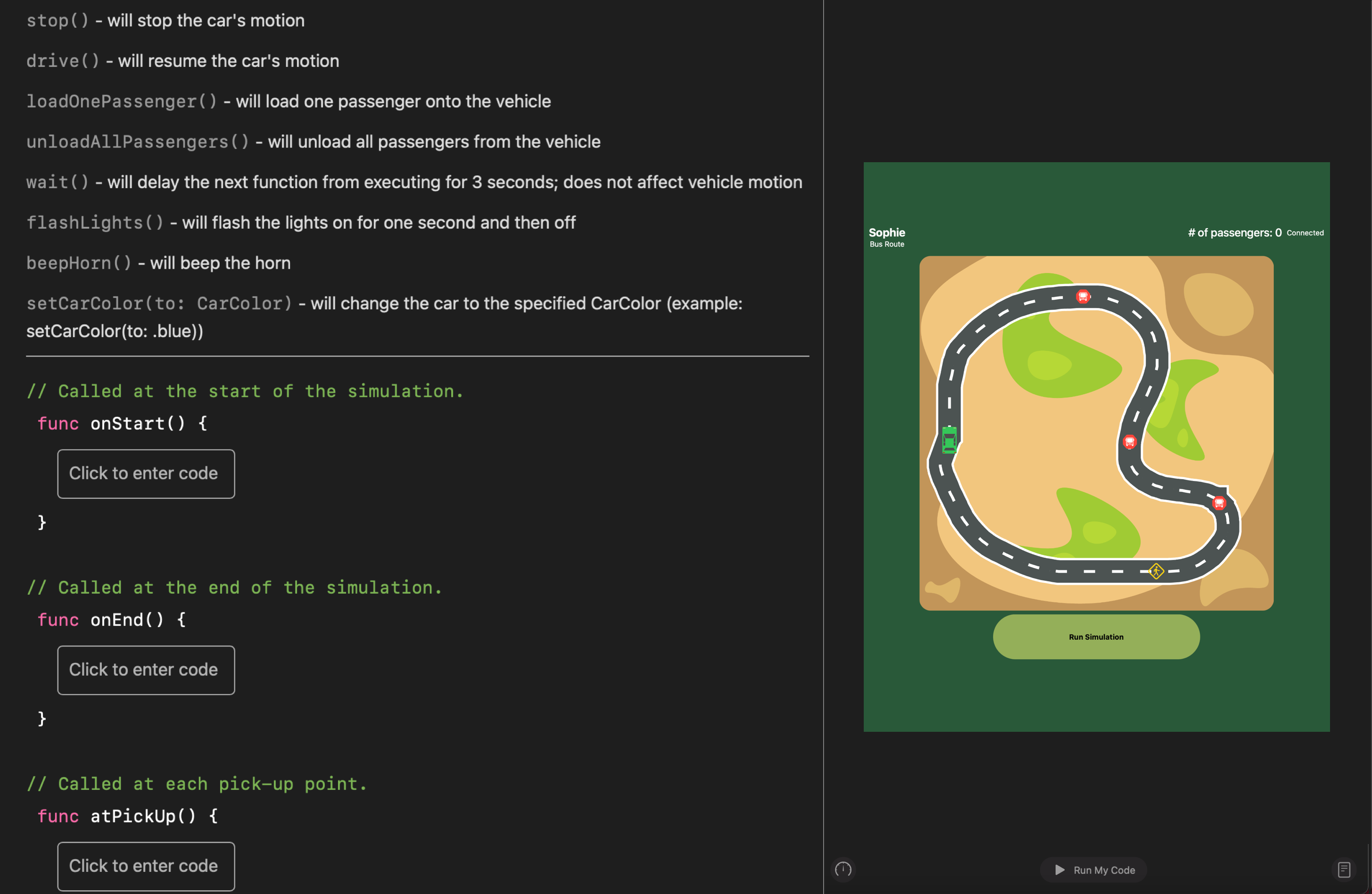}
  \caption{The ARtonomous programming tool's joint model-code interface allows participants to view their objective and function library, write code, and run simulations that call that code all in one place.}
  \label{fig:pgb}
\end{figure*}

\subsection{ARtonomous Model Training Application}
With the ARtonomous model training application, users train RL models for virtual autonomous vehicles through an iPad-based interface that connects to a remote server for model training. 

\subsubsection{iPad Model Creation}
We developed the iPad application in Swift for iOS 14. In the app, students begin by creating a robot (i.e., a model) and giving it a name (e.g., School Bus). From the main track dashboard, students can draw tracks on which they can both test their current robot model and train their robot model for a specified number of episodes (see Figure \ref{fig:training-detail}, Left). Using those tracks, students engage in a process of iterative model improvement by training their robot model on a selected course, testing that robot on the target course, identifying points for improvement based on that testing, and training again on another track to address those key points. Users can also select an individual track to replay simulations of past training episodes, view visualizations of training progress over time (see Figure \ref{fig:training-detail}, Right)---including on-course overlays of the robot's path in prior episodes and endpoints for every episode---or select an episode to see overlay visualizations just for that run. We included these visualizations to aid the development of mental models of the algorithm's decision making processes (e.g., state, action, and reward), and we envisioned the iterative model development process would highlight the impact and role of humans in model training \cite{touretzky2019envisioning, long2020ai}.

\subsubsection{Reinforcement Learning Environment and Model}
Whenever the student elects to train the robot model, the iPad application communicates with a model training server. This server sets the training environment to the track the student selected and runs a reinforcement learning algorithm to update the model, sending training information (e.g., path of the robot and action/reward for each time-step) back to the iPad as episodes are completed.

We created the RL capabilities of the system using an adaptation of the OpenAI car racing environment \cite{brockman2016openai}, with modifications to decrease training time and improve comprehensibility \cite{woodcock2019solving}. This modified environment has a discrete action space with five possible actions (accelerate, brake, left, right, and no change), added functionality for custom tracks, and a new reward function. In addition, we included a simplified scalar grayscale observation space and an expansion to that observation space to include multiple frames (to see velocity and acceleration, not just position), although these changes are purely to speed up model training and are not visible in the user-facing rendering of training or simulations. Model training used the PPO2 algorithm from StableBaselines with a CNN policy, modified to include an episode termination callback that sent episode information via the server back to the iPad application \cite{stable-baselines}.

% \subsubsection{Model Training Server}
% The server provides two distinct services. The first is a web service that authenticates users and allows for storage and access of RL data, including user-drawn tracks and details of training episodes. The second is a RL service, which allows clients to send and receive messages to initiate or update model training and simulation. All data used throughout the the ARtonomous user experience, including data that are generated or consumed within the RL service and the web service were persisted in a database, except for models output by the OpenAI library, which we saved to disk, with paths to the data saved in the database.

\subsection{ARtonomous Programming Tool}
We envision a future of machine learning integrated into robotics education, creating an experience where student activity is a synthesis of programming and machine learning model creation. Therefore, while students use machine learning to train models that drive the robot, they can also practice their coding skills to modify other parts of the robot’s behavior. At any point in the model training process, users can switch to a programming interface to write event-driven code for their bot. This ability to switch back and forth between model creation and programming captures the iterative modeling and programming engineering concepts and practices we aimed to foster, including recognizing and applying the complementary capabilities of ML and programming within robotics systems. 

The ARtonomous programming tool was built for Swift Playgrounds, a development environment for iPadOS and macOS that supports custom tutorials. For example, Lego and Sphero offer Playground ``books'' that allow users to program educational robots like the Lego Mindstorms EV3 and Sphero. We similarly built on top of Swift Playgrounds, developing a model-code interface and function library for ARtonomous using Playgrounds' built-in code compiler and code editor.

This ARtonomous programming tool automatically loads the most up-to-date version of the user's model (retrieving it from the server), allows them to select a course to run their model on, and provides an interface for writing code to run in tandem with the model. 
%When a user runs their code, it calls to the server to run a simulation of the robot navigating the specified course with the current model. 
The models integrate with Swift code via two types of callbacks: built-in callback functions and user-defined waypoints. Built-in callbacks are functions we have included in the Playground Book (\texttt{onStart()}, \texttt{onStep()}, and \texttt{onEnd()}) that are called at their respective times in model simulation execution. These functions allow students to set up or clean up their robot environment (e.g., set the color of the car at the start of the run) or run repeating or time-specific commands (e.g., flash lights on and off at every step). User-defined waypoints, on the other hand, allow users to specify specific locations on the course and then run a callback whenever their agent reaches one of these waypoints. Finally, when writing callback functions, users have access to a custom function library we have created for these virtual robots. Using this library, students can change the car's color, beep the horn, flash its lights, load or unload passengers, pause and resume driving, and more.

%% file: 05-method.tex
\section{Evaluation}
We conducted a study to assess how well ARtonomous achieves our learning and design objectives. Since this research occurred during the COVID-19 pandemic, we conducted testing remotely to protect the health of our participants. Therefore, using a virtual robotics experience, we sought to answer the following research questions pertaining to students' developing understanding of RL, engagement with the system, and curiosity about machine learning more broadly:

\begin{enumerate}
    \item \textbf{RQ1-Developing Understanding} Does this virtual robotics experience help students develop an understanding of reinforcement learning?
    \item \textbf{RQ2-Developing Understanding} How do students using this system learn to identify problems with individual models and generate ideas for how to fix them?
    \item \textbf{RQ3-Developing Understanding} Do the joint modeling and coding capabilities build intuition around which problems in robotics might be solved with code and which might be solved with reinforcement learning?
    \item \textbf{RQ4-Engagement} How engaging do students find the experience of using ARtonomous?
    \item \textbf{RQ5-Curiosity} After using ARtonomous, do students want to learn more about machine learning or reinforcement learning?
\end{enumerate}

\subsection{Participants}
Two researchers conducted this study remotely via video conferencing software with 15 participants aged 11--14 ($M=12.73$, $SD=1.03$; 6 female, 9 male). The researchers followed the same script and alternated roles of facilitator and note-taker between participants. Each session lasted 90 minutes. Participants were recruited from a school for students with reading-related learning differences ($N=1$), a hardware and robotics camp in Colorado ($N=2$), a lower-income middle school in Southern California ($N=5$), and via an email list to employees at a large tech company ($N=7$). We chose these sites to recruit participants with a range of geographic, economic, and personal backgrounds, while also deliberately recruiting from schools/camps with robotics programs or mailing lists for families involved in robotics. To be more inclusive of participants who did not have iPad ($N=5$) and/or Mac ($N=10$) hardware at home, while also socially distancing due to the COVID-19 pandemic, we used screensharing and remote control features to provide virtual access to the hardware as needed. That is, we shared an iPad simulator or a Mac screen with participants who did not have those devices at home, and gave them remote control over our mouse and keyboard to interact with the interfaces directly. In this way, we were able to remotely run this study for participants on a computer running any operating system, regardless of whether they owned an iPad, without compromising the experimental procedure.

All participants had prior experience with programming and exposure to hardware/robotics (e.g., via FIRST Lego League or in-school robotics classes). When asked what they thought machine learning was at the start of the study, 40.0\% talked about a machine getting better on its own, but only 33.3\% discussed the use of experiences or data to drive this improvement, and just one participant (6.7\%) mentioned the machine's ability to approach novel situations. That is, the majority of participants did not know what machine learning was, and those that did had a limited understanding.

\subsection{Objective}
In the study, we gave participants an objective in which they needed to get a School Bus robot to navigate its route (a prescribed course called Bus Route), stop at three bus stops to pick up passengers, and then stop again at the school to drop all of those passengers off. Therefore, using RL, participants trained a virtual robot to navigate around the bus route course. Then, in the programming interface, they handled events to turn the car yellow at the start (\texttt{onStart()}); stop, flash lights, and pick up a passenger at each bus stop (\texttt{atPickup()}); and stop, flash lights, wait, and unload all passengers at the school (\texttt{atDropoff()}).

\subsection{Procedure}
Every participant completed a four-step study procedure, which included phases for introduction, RL model training, programming, and reflection.

\begin{figure*}
  \centering
  \includegraphics[width=\linewidth]{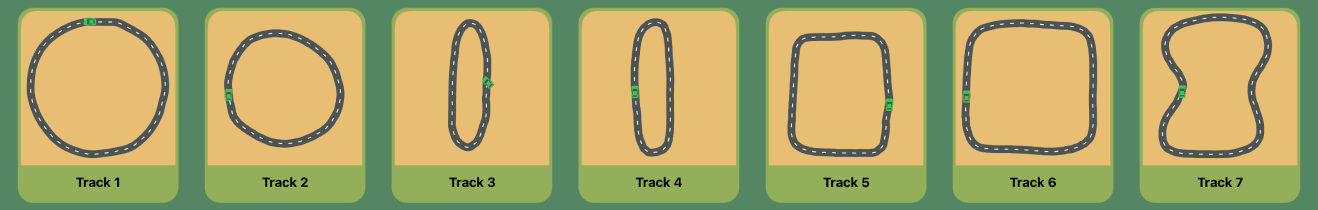}
  \caption{The seven rapid training tracks a participant could choose from during the study. Visually similar tracks (e.g., Track 1 and Track 2) go in opposite directions.}
  \label{fig:rapid-tracks}
\end{figure*}

\subsubsection{Introduction}
During the introduction phase, we first walked participants through the app setup process. For those who owned iPads, we guided them through installing ARtonomous's RL training application via iOS Ad-Hoc distribution (an over-the-air install triggered by scanning a QR code shared via video conferencing screenshare) and, for those who owned a Mac, we guided them through installing Swift Playgrounds and downloading our Playground Book onto their machines. Although the Playground book works on an iPad as well, we chose to have participants use the Playground Book on a computer to ease typing out code and standardize the experience for all participants.

After ensuring the participants could use both of the applications, whether locally or via screenshare, we collected information through verbal interviewing on their programming backgrounds and existing understanding of machine learning. We then showed a 5 minute introductory animation to explain the basics of RL, first via a dog training example followed by an explanation of how these concepts might apply to robot navigation. Finally, we introduced participants to the school bus objective.

\subsubsection{Reinforcement Learning Model Training}\label{study_tasks}
Upon opening the RLbot model training app, the experimenter briefly described the starting state of the robot and the constraints for the study. Specifically, the starting robot was partially pretrained but still not very good at navigating, and the participant was permitted to test their robot (i.e., run a simulation with the current model without training it) as many times as they wanted, but could only train the robot three times.

The experimenter then provided the participant a series of tasks to complete, with necessary prompting along the way to keep the participant on track:
\begin{enumerate}
    \item Recreate the bus route track from the one shown in the objective (see right side of Figure \ref{fig:pgb})
    \item Test your robot on the bus route track and identify any problems it may still have
    \item Create a training track to help your robot gain experience to address that problem
    \item Train your robot on that track
    \item Describe what your robot learned (or didn't learn) based on that training
\end{enumerate}
Participants repeated steps 2--5 three times, completing this process once for each of their training opportunities. They were permitted to ask questions at any time.

Critically, to address the long (i.e., multi-hour) duration of RL model training in conjunction with the constrained timespan of the study, when participants initiated training at step 4, they were prompted to select a track from a list of seven existing ``rapid training'' tracks (see Figure \ref{fig:rapid-tracks}) and asked to explain the reasoning behind their decision. The first six of these tracks were selected because they each target a specific kind of turn the participant might want to direct training toward, whereas Track 7 serves as a more general-purpose training track. We told participants we could train very quickly on these seven tracks and would train just the last ten episodes on the track they had created. In reality, we had pre-trained models for all permutations of these seven tracks; we would load up the appropriate model and run 10 simulations to display as training replays using this next model. In this way, participants could see the progression of training a model even within the time of the user study session and we could gain a deeper understanding of what specific features participants were hoping to train their models on based on their rapid training track choice and subsequent discussion about that choice. Of course, in practice we imagine a user could switch back and forth between model development and coding while training occurred, or set training to run overnight or between class periods.

\subsubsection{Event Handling Code}
After three rounds of training, participants moved over to the ARtonomous programming interface to write event handling code. Here we loaded up the target Bus Route track and its requisite bus stops, reminded participants of their objective, and pointed out the function library before allowing them to jump into programming. The experimenter was able to answer any questions the participant asked and could prompt them to re-read the objective or sections thereof, but did not otherwise assist with the coding.

\subsubsection{Reflection and Feedback}
Participants concluded the study with about 10 minutes of reflection and feedback, described in detail in section \ref{metrics}.

\subsection{Metrics and Collected Data}\label{metrics}
Through both qualitative description and quantitative metrics, we evaluated participants' developing understanding of RL, their engagement with the ARtonomous experience, and their curiosity to learn more. We collected saved track and model data for all participants, recorded video and audio of the study sessions, and transcribed answers to the questions asked in the introduction and reflection portions of the study. We asked all questions to participants verbally, while also screensharing a slide with the written question for the participant to read and refer to. Three researchers collaboratively developed the coding schema used to analyze these quotes and together scored 20\% of participants, discussing and resolving differences. The first author then independently scored the remaining participants according to the agreed upon scoring system.

\subsubsection{RQ1-Developing Understanding: What is RL} 
To examine if participants developed an understanding of reinforcement learning through this experience, we evaluated their knowledge at the end of the session. Similar to evaluative methodology in prior work on educational technology \cite{dietz2021storycoder}, we asked participants to describe what reinforcement learning is, how they used reinforcement learning during the session, and to provide a novel example of reinforcement learning. We looked for key response elements for each answer, and we report percentages of participants who included each element alongside representative quotes to lend further insight. 

% \subsubsection{RQ2-Developing Understanding: Decision Case Study}
% To examine participants' developing understanding of reinforcement learning and the ways in which they use this system to identify and solve problems with individual models, we look to the training decisions they make over time. Specifically, what problems do participants identify in their models, what training tracks do they create to address those problems, and why do they select the rapid training tracks that they do? Through discussion with participants at each decision point, we gain an understanding of how they are reasoning about reinforcement learning at each time-point and how their understanding evolves with new information. We present a representative case study of a single participant's decisions, and discuss it in the context or the broader participant pool.

\subsubsection{RQ2-Developing Understanding: New Track Scenario}
To see if and how participants developed an ability to identify and solve issues with individual models, as part of the reflection and feedback section we presented them with a novel track scenario and a failing robot. This robot successfully followed a wide right turn before failing to turn when that wide right turn transitioned into a very tight left turn. We asked the participants what kind of training track they would create to train this robot; we were looking for participants to specify a tight turn, a left turn, or a transition from a right turn to a left turn.  We report on the content of these responses in terms of percentage of participants that included each correct element. 

\subsubsection{RQ3-Developing Understanding: RL or Code Scenarios}
To probe participants' understandings of the strengths and limitations of RL, we provided them with a series of eight scenarios in robotics (e.g., stop at a stop sign or navigate around an unexpected obstacle) and asked them if the scenario is better solved with RL or programming. We scored their responses against our answer key; four of the scenarios were intended to be solved with RL and four with code.

\subsubsection{RQ4-Engagement}
To measure participants' engagement with the ARtonomous experience, we used the Giggle Gauge, a self-report engagement metric developed for children \cite{dietz2020giggle}. This metric was originally created for use with a bifurcated 4-point scale appropriate to the cognitive limits of children as young as four years old. However, given our older participants, we used the same prompts with a 7-point Likert-type response to allow for more gradation in answers.

\subsubsection{RQ5-Curiosity: Content Analysis of Questions}
To qualitatively describe participants' curiosity about RL after this experience, we showed them a short video of a self-driving car driving along surface streets and a highway. We then asked, ``What questions would you like to ask the creator of this self-driving car?'' We report a brief content analysis of these questions to understand where participants' curiosities lie.

\subsubsection{RQ5-Curiosity: Desire to Learn More}
We quantitatively measured participants' curiosity about RL after this experience by asking them to respond on a 7-point scale the extent to which they would like to learn more about RL.

\subsection{Limitations in Study Design}
We note and explain one deliberate choice that contributes to a limitation in this study design: the constrained time. We chose to limit our study to 90 minutes because we imagine long term usage of this system would accompany more complex design and programming tasks; students might kick off model training and then switch over to programming while waiting for that training to finish. That is, we imagine a child using this system in a classroom setting might interact directly with the model creation interface for short spurts, interspersed with other learning activities (e.g., using the programming interface). Our research questions, though, did not focus heavily on the coding portions of this app. Therefore, we chose to forego a complex coding task that would have further limited our recruitment to students with greater familiarity with Swift. Furthermore, implementing rapid training and condensing these separate interactions into a single experience simplified recruitment and remote study participation for both researchers and for participating families (e.g., for parents who had to give up a work computer for their child to call into the session).

%% file: 06-results.tex
\section{Results}
We report findings regarding understanding of RL, engagement, and curiosity across all 15 participants. All participants completed all three rounds of model training and wrote code that met the objective.

\subsection{RQ1-What is RL}
Participants reported a definition of RL, the way they used RL in the experience, and a novel example of RL. While only 13.3\% ($n=2$) of participants explained that RL was a form of ML (perhaps perceived as an out-of-scope response given the context of the question), 60.0\% ($n=9$) described an agent getting an input, 73.3\% ($n=11$) described an agent choosing an action and receiving rewards or penalties, and 93.3\% ($n=14$) discussed an agent that learned from rewards, penalties, or experiences for future new situations. Participant 11 was one of 6 participants who included all of inputs, actions, rewards, and improvements in their response:  

\begin{quote}
    ``It's when you have some something do something. And it takes your input, and it does an action, and then you give it like---like points, or not points or whatever, um, so that it knows when it did something wrong and when it did something right. So that it can try to do the right thing next time.'' \textit{--P11, age 12}
\end{quote}

When considering how they used RL in the activity, 86.7\% ($n=13$) of participants explained that they used it to train a robot to navigate, but only 40.0\% ($n=6$) spoke about creating targeted training tracks to help it learn and just 20\% ($n=3$) discussed identifying problems in the models, the robot receiving rewards or penalties, or the robot's improvement based on the training. Interestingly, though, 26.7\% ($n=4$) of the participants mentioned their own learning as a form of reinforcement learning. For instance, they described that during the process of coding they could run their code to see when the robot did ``good things'' or ``bad things'' and use that to adjust their code for the next run:

\begin{quote}
    ``I had to look at the track to see if it was doing what I was supposed to do in the objective. And then I used what it did on the track the first time...to change or keep the run in the second time. And then in the end I ended up doing what the objective wanted me to do.'' \textit{--P5, age 13}
\end{quote}

Finally, when describing a new example of reinforcement learning, 93.3\% ($n=14$) participants provided a novel example about learning some kind of behavior or material, 66.7\% ($n=10$) explicitly discussed a reward or penalty, and 73.3\% ($n=11$) explained how gaining experiences might change future outcomes.

\begin{quote}
    ``If you...don't do your homework one day then your teacher punishes you, then the next day you're going to do your homework because the teacher punished you for not doing your homework.'' \textit{--P15, age 13}
\end{quote}

\subsection{RQ2-New Track Scenario}
All participants described one of the three reasonable training tracks for the failing robot, with 86.7\% ($n=13$) describing a tight turn and 26.7\% ($n=4$) suggesting a left turn track (note: some participants included both descriptions in their answer). Additionally 13.3\% ($n=2$) specified that the track should go from one turn into another to specifically target the failing scenario of not transitioning from a right turn to a left turn.

\subsection{RQ3-RL or Code Scenarios}
Participants were successfully able to distinguish which robotics scenarios might be best solved with reinforcement learning and which with code. Out of eight scenario prompts, participants on average answered 85\% correctly (SD=18\%). A single-tailed t-test confirms these scores are significantly above chance, $t(15)=7.90, p<0.001$. 

% Notably, participants were most often marked incorrect on the ``shoot a ball at a detected target'' scenario ($n=5$ marked incorrect). While we intended for this scenario to be solved with code, several participants said it should be done with reinforcement learning so that the robot could train itself to have better aim. That is, even in cases when participants gave an "incorrect" response, their reasoning often demonstrated comprehension of the material.

\subsection{RQ4-Engagement}
We evaluated participant engagement with the ARtonomous experience using a modification of the Giggle Gauge \cite{dietz2020giggle} with a 7-point Likert-type response. Overall we found reasonably high levels of engagement ($M=5.99, SD=1.02$), with all items scoring above 6 on average except aesthetics ($M=5.63, SD=1.23$) and challenge ($M=5.53, SD=1.19$). Notably, if we drop data from one participant who experienced an early bug that led to a multi-second UI delays, we see average engagement increases to 6.17 ($SD=0.77$).

\subsection{RQ5-Inspired Curiosity}
\subsubsection{Content Analysis of Questions}
We categorize the questions participants would want to ask the the creator of a self-driving car into questions about RL, questions about broader machine learning, and non-ML related questions. Most questions fall into the first two categories, indicating a curiosity about machine learning and its applications. Specifically, 40\% ($n=6$) of the participants asked about reinforcement learning (e.g., How many episodes did it take to drive so smoothly?) and another 40\% ($n=6$) about machine learning. Half of machine learning questions related to computer vision or object detection (e.g., How do you program your camera to recognize objects nearby?) and half related to machine decision making (e.g., How does the car deal with something unexpected on the road, like if there's a big rock?). Of the remaining participants, 6.7\% ($n=1$) asked about user experiences/design (i.e., Why is there a steering wheel if the car drives itself?), 6.7\% ($n=1$) listed categories they might discuss (i.e., the complexity of real life compared to simulation) without directly stating a question, and 6.7\% ($n=1$) stated that they had no questions.

\subsubsection{Desire to Learn More}
When asked on a scale from one to seven if they would like to learn more about reinforcement learning, participants unanimously reported that they would. All participants responded with a score of 5 or higher, for a mean score of 6.47 ($SD=0.74$).

%% file: 07-discussion.tex
\section{Discussion}
We began this project with concerns about how the gap between state-of-the-art robotics and educational robotics widens with advancements in machine learning. Further, with educational robotics placing a heavy emphasis on mechanical engineering, the high cost of robotics kits designed for youth can present barriers to participation. By addressing cost and constraints with virtual robots and by building upon new machine learning based conceptual infrastructures, ARtonomous is a more financially and physically accessible educational robotics tool reflective of the types of data-driven approaches used in professional robotics today.

Our evaluation demonstrates that participants were able to develop an understanding of reinforcement learning through the ARtonomous experience (i.e., what it is, how it fails, and when it applies), that they were highly engaged with the application, and that they were curious to learn more about reinforcement learning and its real-word uses. All participants could suggest reasonable training tracks for a novel track scenario and were able to---at least in part---explain reinforcement learning, its usage, and an example thereof. Participants reported high levels of engagement with the app and an even higher desire to learn more about the subject material. Collectively, these results demonstrate ARtonomous' success in addressing our key research objectives: 1) creating low cost, high engagement robotics education tools that help learners explore reinforcement learning and 2) bridging the gap between educational and professional robotics to develop student understanding of reinforcement learning, support positive engagement with the content, and build student curiosity toward machine learning more broadly.

\subsection{Low-Cost and High Engagement with Virtual Robotics}
Surveying the landscape of educational robotics tools, the cost of specialized hardware can inhibit broad reach. In the United States in particular, economic disadvantage is strongly correlated with race and ethnicity. Consequently, expensive tools can contribute to the maintenance of racist barriers to educational equality. 

These concerns validate core premises of the ARtonomous vision: that we can use general purpose computing devices to simulate robotic systems and visualize interaction between those systems and the surrounding environment. Our research illustrates that training and programming simulated robots is engaging and educational for participants. This result is consistent with prior work on virtual robotics \cite{berland2015comparing, eguchi2012student, witherspoon2017developing}, with the additional inclusion of ML model training.

\begin{figure}
  \centering
  \includegraphics[width=0.95\linewidth]{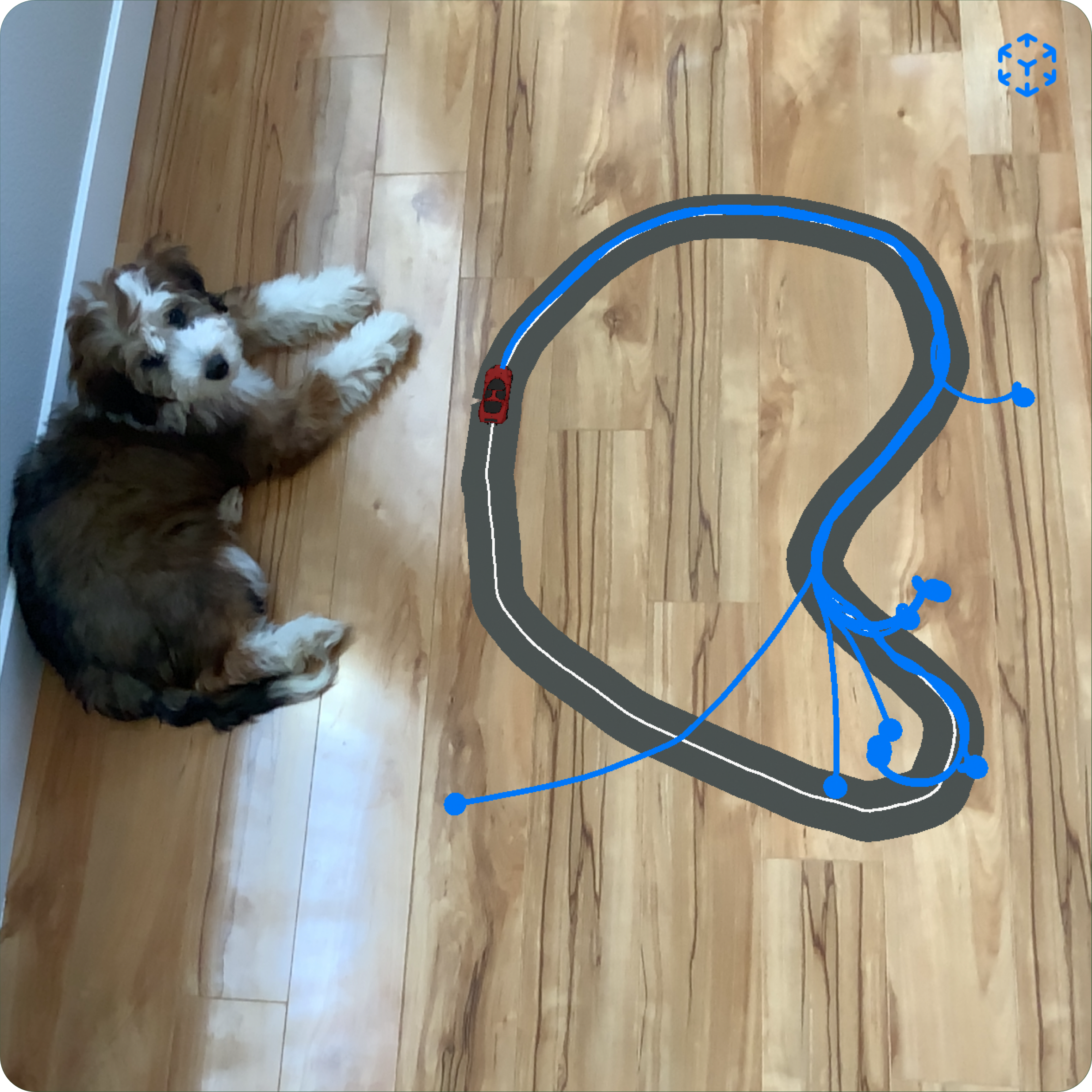}
  \caption{Prototype augmented reality features allow users to view their track, robot, and training overlay details in physical space.}
  \label{fig:ar}
\end{figure}

Looking ahead, we foresee that augmented reality (AR) functionality has the potential to take this engagement and learning even further \cite{dede2009immersive}, still without requiring specialized (and potentially cost-prohibitive) hardware. We built prototype functionality of AR integration that allows robot training---including replays, test simulations, and overlay visualizations---to appear in the physical-virtual space (see Figure \ref{fig:ar}). Although we chose to forego testing of the AR components within this tool, instead prioritizing the geographic and economic diversity of our participant pool, a large body of prior work shows that AR educational experiences can improve learning and engagement \cite{dede2009immersive, radu2012should}. We look forward to future work exploring AR's potential in robotics education contexts. 

\subsection{Bridging the Gap Between Educational and Professional Robotics}
All fields of disciplined human endeavor involve representational infrastructures---tools that participants in those fields use to frame, communicate about, and accomplish work \cite{hall2002disrupting,suchman1988representing}. Learning a field, whether robotics, archaeology, or policing \cite{goodwin1994professional}, is fundamentally about learning to see and do as those already within that field do. As we discuss above, the rapid advancement of ML in professional robotics combined with the relative absence of ML in educational robotics is progressively relegating educational robotics into a less and less authentic enterprise. The representational infrastructure of educational robotics is generally lacking methods of ML, including RL, and norms that emphasize the contemporary importance of autonomy. Our findings show 1) that RL (and ML) can be a part of how students learn robotics and 2) that the representational infrastructure gap between robotics and educational robotics is closable via new systems like the one we present here.

% At present, many robotics education tools and programs place a heavy emphasis on mechanical engineering. For instance, the FIRST LEGO League requires use of the LEGO SPIKE Prime or EV3 kits, which include 528 and 601 pieces, respectively. Participants use these pieces to assemble robotic vehicles, including  mechanical mechanisms to connect motors to wheels, arms, and other actuators. Learning to assemble these pieces in robust and functional ways---in essence, learning mechanical engineering---is \textit{the} central technical challenge of programs like FIRST \cite{first_2021}. Our needfinding interviews showed that teachers sometimes found it overwhelming to try to facilitate both  mechanical engineering and computing, and so chose to pick one or the other. Given our goal of furthering the integration of ML into robotics education (and computing education more broadly), we investigate educational futures that would encourage the inclusion of computing. 

Our research illustrates how new systems can foster approaches to teaching and learning that place more emphasis on the computer science of ML and programming. The iPad-based ARtonomous experience we evaluated involves several representational practices that are lacking or uncommon in today's most common educational robotics platforms and experiences, including creating idealized contexts (i.e., training and simulation environments) for robot operation, identifying specific examples of problematic or successful robotic behavior from simulations within those contexts, noting patterns in aggregated replays of multiple simulations, drawing focused training scenarios to address problematic patterns, evaluating ML model improvement over time, juxtaposing code with simulations of RL-based model execution, and observing and debugging the execution of programs that are combinations of RL models and user-defined code.

Some of these representational practices have analogues in today's physical robotics education experiences. For example, participants might sketch arenas that they wish for their robots to execute within and then draw the trajectories that they want their robots to follow within those arenas. They might video record their robots in action and replay those later in order to share their successes or document opportunities for improvement. Virtual robotics platforms, such as the Virtual Robotics Toolkit \cite{vrt}, also afford the ability to capture and replay simulations.

However, the essence of this work---iteratively training an ML-based navigation model using a series of training environments designed in light of past model performance and then integrating the ML model with user-authored code---involves a set of practices that are not available in other tools but are reflective of broader AI education objectives \cite{touretzky2019envisioning, long2020ai}. Moreover, ML could enable young people to create robotic systems that are currently difficult, if not impossible, for them to do with programming alone (e.g., create autonomous vehicles that navigate model cities without pre-programmed routes). We further envision future RL education platforms for middle school learners that introduce additional aspects of model training (e.g., observation space, exploration rate, or reward function) in similarly engaging and accessible ways. The approachability of ML practices, as evidenced by our findings, raises an essential question for robotics education: What are the most important learning goals for robotics education in a world where authentic robotics practice involves a data-driven synthesis of ML and programming in order to create autonomous systems? This work suggests that the skills of creating, assessing, and integrating ML models should and could be among those goals. 

%% file: 08-conclusion.tex
\section{Conclusion}
We presented ARtonomous, a tablet-based application that introduces middle school students to reinforcement learning through virtual autonomous robotics. We address cost constraints in educational robotics while still maintaining high engagement by leveraging a virtual robot approach. Further, we aim to close the gap between educational and professional robotics through the introduction of machine learning techniques. Through a 90-minute evaluation study with middle school aged students (ages 11--14) we demonstrated the efficacy of ARtonomous in achieving these design goals. Specifically, ARtonomous supported participants in developing an understanding of reinforcement learning, led to high-levels of self-reported engagement, and inspired curiosity among participants for further learning about RL. Having demonstrated that RL can be made accessible and approachable for middle school students and that it can integrate into robotics education, our work highlights opportunities for robotics educators, HCI and education researchers, and product developers to reassess the most important learning goals in educational robotics and to consider making space for RL or ML via virtual simulation methods.

%% file: 09-selection-of-children.tex
\section{Selection and Participation of Children}
Youth were recruited from a school for students with reading-related learning differences, a hardware and robotics camp in Colorado, a middle school in Southern California serving a lower-income community, and via an email list to employees at a large technology company. We chose these sites to recruit participants with a range of geographic, economic, and personal backgrounds. All youth provided verbal assent to participate in the study and to be video recorded, and youth and parents signed consent forms prior to video and audio data collection. Families were told they could choose to end the study at any time and still receive a gift card; families received a \$25 gift card in exchange for participating.